\documentclass[onecolumn,draftclsnofoot,12pt]{IEEEtran}
\usepackage{ifpdf}
\usepackage[nospace]{cite}
\usepackage[cmex10]{amsmath}
\usepackage{array}
\usepackage{amsfonts}
\usepackage{mathrsfs}
\usepackage{arydshln}
\usepackage{slashbox}
\usepackage{graphicx}
\usepackage{float}
\usepackage{subfigure}
\usepackage{amssymb}
\usepackage{amsmath}
\usepackage[colorlinks,linkcolor=black,anchorcolor=black,citecolor=black,hyperfootnotes=true]{hyperref}
\usepackage{multirow}
\usepackage{multicol}
\usepackage{cite}
\usepackage[subfigure]{graphfig}
\usepackage{xcolor}
\usepackage{setspace}

\usepackage[linesnumbered,ruled]{algorithm2e}

\newtheorem{proposition}{\underline{Proposition}}

\newcommand{\qed}{\nobreak \ifvmode \relax \else
      \hskip1.5em plus0em minus0.5em \fi \nobreak
      \vrule height0.75em width0.5em depth0.25em\fi}

\begin{document}
\title{Joint Altitude and Beamwidth Optimization for UAV-Enabled Multiuser Communications \thanks{The authors are with the Department of Electrical and Computer Engineering, National University of Singapore (e-mail: \{haiyun.he,shuowen.zhang\}@u.nus.edu, \{elezeng,elezhang\}@nus.edu.sg).}
\author{\IEEEauthorblockN{Haiyun~He, Shuowen~Zhang,~\IEEEmembership{Student Member,~IEEE}, Yong~Zeng,~\IEEEmembership{Member,~IEEE}, Rui~Zhang,~\IEEEmembership{Fellow,~IEEE}}}}

\maketitle

\begin{abstract}
In this letter, we study multiuser communication systems enabled by an unmanned aerial vehicle (UAV) that is equipped with a directional antenna of adjustable beamwidth. We propose a \emph{fly-hover-and-communicate} protocol where the ground terminals (GTs) are partitioned into disjoint clusters that are sequentially served by the UAV as it hovers above the corresponding cluster centers. We jointly optimize the UAV's flying altitude and antenna beamwidth for throughput optimization in three fundamental multiuser communication models, namely UAV-enabled downlink multicasting (MC), downlink broadcasting (BC), and uplink multiple access (MAC). Our results show that the optimal UAV altitude and antenna beamwidth critically depend on the communication model considered.
\end{abstract}
\begin{IEEEkeywords}
UAV communication, altitude optimization, directional antenna, beamwidth optimization, wireless network.
\end{IEEEkeywords}
\section{Introduction}
Wireless communication assisted by unmanned aerial vehicles (UAVs) is a promising technology to meet the highly diversified and dynamic data demands in future wireless systems \cite{649}. Compared to existing technologies such as small cell and satellite communication, UAV-enabled wireless communication has appealing advantages, such as the ability of on-demand and fast deployment, higher capacity due to dominant line-of-sight (LoS) communication links with the ground terminals (GTs), and additional design degrees of freedom by exploiting the fully controllable UAV mobility. Thus, UAV-enabled communication is expected to play a significant role in future wireless systems, especially for applications such as data offloading for cellular base stations (BSs) in temporary hotspot areas, mobile relaying for emergency responses, periodic information dissemination and data collection in large Internet of Things (IoT) networks, etc.

To realize the full potential of UAV-enabled communication, it is crucial to maximally exploit the fully controllable UAV mobility in the three-dimensional (3D) space. The horizontal and/or vertical positions of the UAVs could be optimized for their deployment, leading to various two-dimensional (2D) or 3D UAV placement designs \cite{642,793,803,UAV_D2D,sharma2016uav,886,Lyu2016Cyclical}. Furthermore, the UAVs' locations could be contiguously adjusted over time to best meet the communication requirement, which leads to the more general UAV trajectory optimization problems \cite{641}, \cite{904}.

However, the existing works mainly assume that the UAVs are equipped with either omnidirectional antenna or directional antenna with fixed beamwidth. In this letter, we consider the new case where the UAV is equipped with a directional antenna whose beamwidth can be adjusted. Note that with contemporary beamwidth tuning technologies \cite{906}, antennas with tunable radiation patterns have already been applied for various applications such as satellite communication and remote sensing. For UAV-enabled wireless communication systems with tunable antenna beamwidth, there is in general an interesting trade-off in adjusting the antenna beamwidth of the UAV versus its altitude above a ground position. Specifically, for a given UAV altitude, increasing the antenna beamwidth helps cover more GTs within the antenna's main lobe, but at the cost of reduced link capacity for each of those GTs due to the reduced antenna gain in the main lobe. On the other hand, for a given antenna beamwidth, an increase in the UAV altitude would cover more GTs, but with lower link capacity for each GT due to the increased link distance.

To optimally resolve such a trade-off, we study in this letter the joint UAV altitude and beamwidth optimization problem for UAV-enabled multiuser communication systems. We propose a practical {\it fly-hover-and-communicate} protocol, where the GTs are partitioned into disjoint clusters with the size of each cluster determined by the area projected by the antenna's main lobe on the ground. Each cluster is then sequentially served by the UAV as it hovers above the corresponding cluster center. Note that although this protocol may be suboptimal in general, it is favorable for practical implementation.\footnote{In this letter, we assume that \emph{rotary-wing} UAV is adopted, which is capable of hovering at desired locations. Moreover, we assume delay-tolerant communications between the UAV and the GTs (e.g., periodic sensing for the uplink and information dissemination in a public safety network for the downlink), where the tolerable delay for each GT is sufficiently large such that any delay incurred in our proposed design does not affect the quality-of-service (QoS) at the GTs.} We study three fundamental UAV-enabled multiuser communication models, namely, {\it downlink multicasting  (MC)}, where the UAV sends common information to all GTs in each cluster, {\it downlink broadcasting} (BC), where the UAV sends independent information to different GTs via frequency division multiple access (FDMA), and {\it uplink multiple access (MAC)}, where each GT sends independent information to the UAV via FDMA. Our results show that the optimal UAV altitude and antenna beamwidth critically depend on the communication model considered. Specifically, in terms of the UAV altitude, it should be set as the maximum possible value for downlink MC, but the minimum possible value for downlink BC, while it can be any feasible value for uplink MAC since the throughput is shown to be independent of the UAV altitude. On the other hand, for antenna beamwidth, an optimal value exists for downlink MC, while it should be set to the minimum feasible value for downlink BC, and its effect on uplink MAC is shown to be marginal.
\section{System Model}\label{sec_sys}
\begin{figure}[t]
	\centering
    \includegraphics[width=12cm]{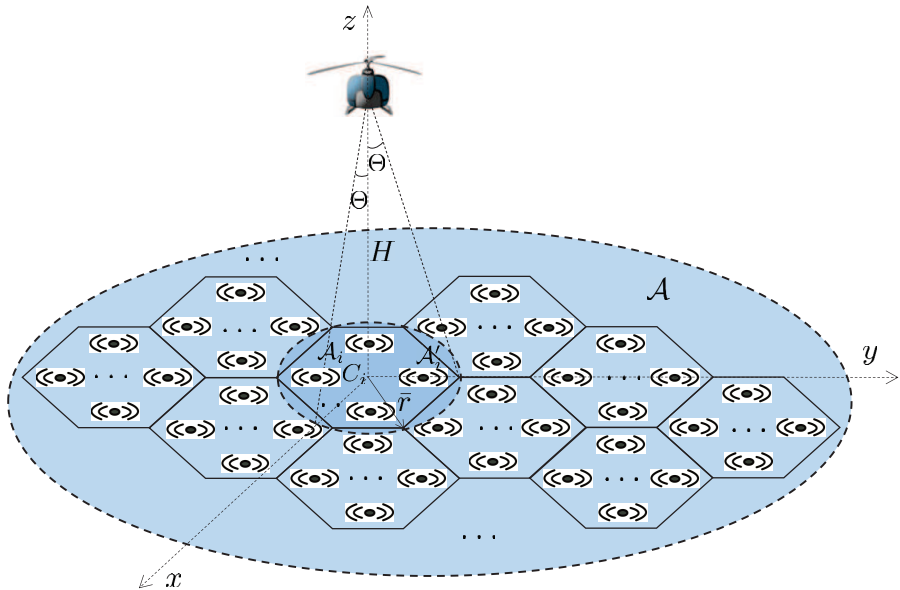}
    \caption{A UAV-enabled outdoor wireless communication system.}\label{system_diagram}
\end{figure}

We consider a UAV-enabled wireless communication system as shown in Fig. \ref{system_diagram}, where the UAV is deployed as a flying BS at an altitude of $H$ meters (m) to serve $K$ GTs in a large area $\mathcal{A}$ of size $A$ $\hbox{m}^2$. The GTs are assumed to be uniformly distributed in $\mathcal{A}$ with density $\rho=\frac{K}{A}$ GTs/$\mbox{m}^2$. We assume that the UAV is equipped with a directional antenna of adjustable beamwidth. For simplicity, we assume that the azimuth and elevation half-power beamwidths of the UAV antenna are equal, which are both denoted as $2\Theta$ in radians (rad), with $\Theta\in \left(0,\frac{\pi}{2}\right)$. Moreover, the corresponding antenna gain in direction $(\theta,\psi)$ is approximately modelled as
\begin{equation}\label{AG}
G=\begin{cases}
\frac{G_{0}}{\Theta^2},\quad &-\Theta\leq \theta \leq \Theta, -\Theta\leq \psi \leq \Theta\\
g\approx 0,\quad &\mbox{otherwise},
\end{cases}
\end{equation}
where $G_0=\frac{30000}{2^2}\times\left(\frac{\pi}{180}\right)^2\approx2.2846$; $\theta$ and $\psi$ denote the azimuth and elevation angles, respectively (\!\!\cite{903}, Eq. (2-51)). Note that $g$ satisfies $0< g\ll \frac{G_0}{\Theta^2}$ in practice, and we assume $g=0$ for simplicity. On the other hand, we assume that each GT is equipped with an omnidirectional antenna with unit gain. Thus, for any given UAV location, the disk region on the ground that is covered by the antenna's main lobe with radius $\bar{r}=H\tan\Theta$ corresponds to the ground \emph{coverage area} of the UAV, as illustrated in Fig. \ref{system_diagram}. Furthermore, we assume that the GTs are located outdoors in rural areas, and the communication channel between the UAV and each GT is dominated by the LoS path,\footnote{Note that for other scenarios (e.g., where GTs are located indoors and/or in urban areas), the LoS channel model serves as a benchmark and helps to characterize the performance limits of more practical scenarios.} thus the channel power gain between the UAV and a GT with horizontal distance $r\leq \bar{r}$ to the UAV is given by
\begin{equation}\label{hr}
h(r)=\frac{{\beta}_0}{H^2+r^2},
\end{equation}
where ${\beta}_0$ denotes the channel power gain at the reference distance $d_0=1$m.

We assume that the area $\mathcal{A}$ is sufficiently large such that it can be partitioned into $N=\frac{A}{A_s}$ tessellated regular hexagonal cells denoted by $\{\mathcal{A}_i\}_{i=1}^N$ for any given $\bar{r}$, each with equal circumradius $\bar{r}$ and size $A_s=\frac{3\sqrt{3}}{2}\bar{r}^2$, as illustrated in Fig. \ref{system_diagram}. The average number of GTs in each $\mathcal{A}_i$ is then given by
\begin{equation}\label{Ks}
K_s=\rho A_s=\frac{3\sqrt{3}}{2}\rho H^2\tan^2\Theta.
\end{equation}
It is worth noting that the $K_s$ GTs in each $\mathcal{A}_i$ all lie within the UAV coverage area when the UAV is hovering above the center of $\mathcal{A}_i$, denoted as $C_i$. Under the proposed fly-hover-and-communicate protocol, the UAV sequentially serves the GTs in $\mathcal{A}$ by successively flying over $C_i$'s based on a certain order (e.g., that obtained via algorithms for solving the travelling salesman problem to minimize the total flying distance \cite{TSP}), and hovering above each $C_i$ for time duration $T_i>0$ seconds (s) to communicate with the corresponding $K_s$ GTs in $\mathcal{A}_i$. In this letter, we assume that the UAV flying speed $V_m$ is sufficiently large and the sum hovering time is much larger than the UAV flying time, i.e.,  $\sum_{i=1}^N T_i \gg \frac{L}{V_m}$, where $L$ denotes the total traveling distance. Thus the mission completion time can be expressed as $T_{\mathrm{completion}}=\sum_{i=1}^N T_i+\frac{L}{V_m}\approx\sum_{i=1}^N T_i$.

We assume that the feasible range of altitude $H$ of the UAV is given by $[H_{\min},H_{\max}]$, where $H_{\min}>0$ and $H_{\max} > H_{\min}$ are practically determined by e.g., obstacle heights and authority regulations; and the feasible range of half-beamwidth $\Theta$ is assumed to be $\left[\Theta_{\min},\Theta_{\max}\right]$, where $\Theta_{\min}>0$ and $\Theta_{\min}<\Theta_{\max}<\frac{\pi}{2}$ are determined by the practical antenna beamwidth tuning technique adopted (e.g., \cite{906}). For all multiuser systems considered, we denote $W$ as the total communication bandwidth in Hz; $P_d$ in watt (W) as the transmission power at the UAV for the two downlink cases (MC and BC) and $P_u$ as the transmission power by  each GT for the uplink case (MAC). Furthermore, $N_0$ denotes the noise power spectrum density at all receivers in W/Hz.

\section{Downlink Multicasting}\label{sec_multicast}
In this section, we consider the case of downlink MC, where the UAV has a mission to deliver a common file of total size $\bar{D}$ bits to all GTs in $\mathcal{A}$. Our objective is to minimize the mission completion time by jointly optimizing $H$ and $\Theta$.
As clarified in Section \ref{sec_sys}, the mission completion time can be approximated by $T_{\mathrm{MC}}\approx\sum_{i=1}^{N} T_i$.

When the UAV hovers above $C_i$, the received signal-to-noise ratio (SNR) at a GT with distance $r$ to $C_i$ is given by
\begin{equation}
\gamma_{\mathrm{MC}}(r)=\frac{P_dGh(r)}{N_0W}=\frac{\alpha}{\Theta^2(H^2+r^2)},\quad 0\leq r\leq \bar{r},
\end{equation}
where $\alpha=\frac{P_dG_0{\beta}_0}{N_0W}$. The corresponding achievable communication rate is thus given by $R_{\mathrm{MC}}(r)=\log_2(1+\gamma_{\mathrm{MC}}(r))$ in bits per second per Hz (bps/Hz). By noting that $R_{\mathrm{MC}}(r)$ is a decreasing function of $r$, it follows that the common file can be completely delivered to all GTs in $\mathcal{A}_i$ if $T_iWR_{\mathrm{MC}}(\bar{r})\geq \bar{D}$ holds, with $R_{\mathrm{MC}}(\bar{r})$ denoting the achievable rate of the cell-edge GT. Hence, the mission completion time for downlink MC with the proposed scheme can be written as
\begin{equation}\label{TMC}
T_{\mathrm{MC}}=\frac{N \bar{D}}{WR_{\mathrm{MC}}(\bar{r})}=
\frac{\frac{K\bar{D}}{W}}{K_sR_{\mathrm{MC}}(\bar{r})}=
\frac{\frac{K\bar{D}}{W}}{\tilde{R}_{\mathrm{MC}}}.
\end{equation}
Consequently, $T_{\mathrm{MC}}$ is minimized by maximizing $\tilde{R}_{\mathrm{MC}}\overset{\Delta}{=}K_sR_{\mathrm{MC}}(\bar{r})$, which can be explicitly expressed as
\begin{equation}\label{RMC}
\!\!\tilde{R}_{\mathrm{MC}}(H,\Theta)\!=\!\frac{3\sqrt{3}}{2}\rho H^2\tan^2\Theta\log_2\left(1\!+\!\frac{\alpha\cos^2\Theta}{\Theta^2H^2}\right).
\end{equation}

It is worth noting that as $H$ or $\Theta$ increases, $K_s$ given in (\ref{Ks}) increases, while ${R}_{\mathrm{MC}}(\bar{r})$ decreases. To balance the above trade-off and find the optimal $H$ and $\Theta$ for maximizing $\tilde{R}_{\mathrm{MC}}(H,\Theta)$, we present the following results.
\begin{proposition}\label{prop_MC_H}
For any given $\Theta\in[\Theta_{\min},\!\Theta_{\max}]$, $\tilde{R}_{\mathrm{MC}}(H,\Theta)$ is a non-decreasing function of $H$, with $H>0$.
\end{proposition}
\begin{IEEEproof}
Define $\tilde{\alpha}_1=\frac{\alpha\cos^2\Theta}{\Theta^2},\tilde{\alpha}_2=\frac{3\sqrt{3}}{2\ln2}\rho \tan^2\Theta$ and $\tilde{H}=H^2$. It can be shown that $\underset{\tilde{H}\rightarrow\infty}{\lim}\!\frac{\partial \tilde{R}_{\mathrm{MC}}(H,\Theta)}{\partial \tilde{H}}=0$, $\frac{\partial^2 \tilde{R}_{\mathrm{MC}}(H,\Theta)}{\partial \tilde{H}^2}=\frac{-\tilde{\alpha}_1^2\tilde{\alpha}_2}{\tilde{H}(\tilde{\alpha}_1+\tilde{H})^2}<0$, which imply that $\frac{\partial \tilde{R}_{\mathrm{MC}}(H,\Theta)}{\partial \tilde{H}}\geq 0,\forall \tilde{H}>0$.
\end{IEEEproof}

Proposition \ref{prop_MC_H} implies that for any given $\Theta$, the optimal $H$ is $H^\star_{\mathrm{MC}}=H_{\max}$. This is because as $H$ grows, the increase in $K_s$ is more significant than the decrease in ${R}_{\mathrm{MC}}(\bar{r})$.

Next, we investigate the effect of $\Theta$ on $\tilde{R}_{\mathrm{MC}}(H,\Theta)$ with given $H=H_{\max}$. However, due to the complicated expression of the derivative of $\tilde{R}_{\mathrm{MC}}(H,\Theta)$ with respect to $\Theta$, it is generally difficult to obtain a closed-form solution of the optimal $\Theta$ to maximize $\tilde{R}_{\mathrm{MC}}(H,\Theta)$, which is denoted as $\Theta^\star_{\mathrm{MC}}$. As a result, $\Theta^\star_{\mathrm{MC}}$ needs to be obtained via a one-dimensional search over $\left[\Theta_{\min},\Theta_{\max}\right]$, for which the numerical results will be given later in Section \ref{sec_num}. Nevertheless, it can be shown from (\ref{RMC}) that $\underset{H\rightarrow \infty}{\lim}\tilde{R}_{\mathrm{MC}}(H,\Theta)=\frac{3\sqrt{3}}{2}\rho \alpha\frac{\sin^2\Theta}{\Theta^2}$, which yields $\underset{H\rightarrow\infty}{\lim}{\Theta_{\mathrm{MC}}^\star}=\Theta_{\min}$, since $\frac{\sin^2\Theta}{\Theta^2}$ is a decreasing function of $\Theta$, $0<\Theta<\frac{\pi}{2}$. Thus, when $H_{\max}$ is sufficiently large, the corresponding optimal beamwidth is $\Theta_{\mathrm{MC}}^\star=\Theta_{\min}$.
\section{Downlink Broadcasting}\label{sec_broadcast}
Next, we consider the case of downlink BC where the UAV needs to send independent information to each of the GTs in $\mathcal{A}$. Our objective is to maximize the GTs' sum throughput within a given period $T_{\mathrm{BC}}$ (in s) via joint optimization of $H$ and $\Theta$. Let the sum rate of all GTs in each cell $\mathcal{A}_i$ be denoted by $R_{\mathrm{BC}}$ in bps/Hz. Then the total throughput of all cells is given by $D_{\mathrm{BC}}=\sum_{i=1}^N T_iWR_{\mathrm{BC}}\!=\!T_{\mathrm{BC}}WR_{\mathrm{BC}}$ in bits. Hence, maximizing $D_{\mathrm{BC}}$ is equivalent to maximizing $R_{\mathrm{BC}}$. Note that a closed-form expression of $R_{\mathrm{BC}}$ is generally difficult to obtain since it involves integration of \hbox{GT rates over} uniform distribution in the hexagonal cell $\mathcal{A}_i$. For analytical tractability, we assume that when the UAV hovers above each $C_i$, it serves all GTs within the disk region $\mathcal{A}_i'$ centered at $C_i$ with the same radius $\bar{r}$ as $\mathcal{A}_i$, as shown in Fig. \ref{system_diagram}, which is of size $A_s'=\pi \bar{r}^2$. Thus, there are on average $K_s'=\rho A_s'=\rho\pi H^2\tan^2\Theta$ GTs in $\mathcal{A}_i'$, whose sum rate is denoted by $\tilde{R}_{\mathrm{BC}}$ in bps/Hz. In the sequel, we aim to maximize $\tilde{R}_{\mathrm{BC}}$ as an approximation of $R_{\mathrm{BC}}$.

We assume FDMA for the $K_s'$ GTs to be simultaneously served by the UAV in each $\mathcal{A}_i'$, where each GT is allocated with an equal bandwidth $\frac{W}{K_s'}$. Moreover, we consider the equal power allocation scheme, where the total UAV downlink transmission power $P_d$ is equally allocated to the $K_s'$ GTs in each $\mathcal{A}_i'$. Thus, when the UAV hovers above $C_i$, the received SNR at a GT with distance $r$ to $C_i$ is given by
\begin{equation}
\gamma_{\mathrm{BC}}(r)=\frac{\frac{P_{d}}{K_{s}'}G h(r)}{N_{0}\frac{W}{K_{s}'}}
=\frac{\alpha}{\Theta^2(H^2+r^2)},\quad 0\leq r\leq \bar{r}.
\end{equation}
The corresponding GT rate is then given by  ${R}_{\mathrm{BC}}(r)\!=\!\frac{1}{K_{s}'}\log_{2}\left(1+\gamma_{\mathrm{BC}}(r)\right)
=\frac{\log_2\left(1+\frac{\alpha}{\Theta^2(H^2+r^2)}\right)}{\rho{\pi}H^2\tan^2\Theta}$ bps/Hz. Hence, the total communication rate of all GTs in $\mathcal{A}_i'$ is given by
\begin{align}\label{R_BC-E}
&\tilde{R}_{\mathrm{BC}}(H,\Theta)=\int_{0}^{2\pi}\int_{0}^{H\tan\Theta} \rho {R}_{\mathrm{BC}}(r) r dr d\theta\nonumber\\
=&\frac{1}{\sin^2\Theta}\log_{2}\left(1+\frac{\alpha\cos^2\Theta}{\Theta^2 H^2}\right)-\frac{1}{\tan^2\Theta}\log_{2}\left(1+\frac{\alpha}{\Theta^2 H^2}\right)\nonumber\\
+&\frac{\alpha }{\Theta^2H^2\tan^2\Theta}\log_{2}\left(\frac{\Theta^2H^2+\alpha\cos^2\Theta}{\Theta^2H^2\cos^2\Theta+\alpha\cos^2\Theta}\right).
\end{align}

Note that there exists a trade-off between the individual GT rate, ${R}_{\mathrm{BC}}(r)$, and the number of GTs in $\mathcal{A}_i'$, $K_s'$, which are decreasing and increasing functions of both $H$ and $\Theta$, respectively. To obtain the optimal $H$ and $\Theta$ to maximize $\tilde{R}_{\mathrm{BC}}(H,\Theta)$, we provide the following proposition.
\begin{proposition}\label{prop_BC-E_H}
For any given $\Theta\in [\Theta_{\min},\Theta_{\max}]$, $\tilde{R}_{\mathrm{BC}}(H,\Theta)$ is a decreasing function of $H$, with $H>0$.
\end{proposition}
\begin{IEEEproof}
Let $\tilde{H}=H^2$. It can be shown that $\frac{\partial \tilde{R}_{\mathrm{BC}}(H,\Theta)}{\partial \tilde{H}}=\frac{\alpha}{\tilde{H}^2\Theta^2\tan^2\Theta}\bigg(\log_2\bigg({\tilde{H}+\frac{\alpha}{\Theta^2}}\bigg)-\log_2\bigg({\frac{\tilde{H}}{\cos^2\Theta}+\frac{\alpha}{\Theta^2}}\bigg)\bigg)$, which is negative since $\cos^2\Theta< 1$ for any $\Theta\in [\Theta_{\min},\Theta_{\max}]$.
\end{IEEEproof}

Proposition \ref{prop_BC-E_H} indicates that given any $\Theta$, the optimal $H$ in the downlink BC case is ${H_{\mathrm{BC}}}^\star=H_{\min}$, which is in sharp contrast to the previous case of downlink MC in Section \ref{sec_multicast}. This can be intuitively explained as follows. Note that due to the bandwidth partitioning for FDMA, as $H$ increases, each individual GT rate in downlink BC decreases more quickly than that in downlink MC, since in the BC case each GT is assigned smaller portion of the total bandwidth and less power with the increasing number of GTs served. On the other hand, due to the difficulty in deriving a closed-form expression of the optimal $\Theta$ from (\ref{R_BC-E}), we examine the effect of the beamwidth $\Theta$ on $\tilde{R}_{\mathrm{BC}}(H,\Theta)$ numerically in Section \ref{sec_num}.
\section{Uplink Multiple Access}\label{sec_uplink}
Last, we consider the case of uplink MAC, where each GT in $\mathcal{A}$ needs to send independent information to the UAV. Our objective is to maximize the GTs' sum throughput within a given period $T_{\mathrm{MAC}}$ (in s) by jointly optimizing $H$ and $\Theta$. Similar to the case of downlink BC in Section \ref{sec_broadcast}, we assume that the UAV serves all $K_s'$ GTs in $\mathcal{A}_i'$ via FDMA with equal bandwidth allocation over the served GTs, thus the sum throughput is maximized by maximizing the total communication rate of all GTs in $\mathcal{A}_i'$, denoted as $\tilde{R}_{\mathrm{MAC}}$.

When the UAV hovers above $C_i$, the received SNR at the UAV from a GT with distance $r$ to $C_i$ is given by
\begin{equation}\label{gamma_UL}
\gamma_{\mathrm{MAC}}(r)=\frac{P_uh(r)G}{N_{0}\frac{W}{K_{s}'}}=\frac{\eta H^2\tan^2\Theta}{\Theta^2(H^2+r^2)},\quad 0\leq r\leq \bar{r},
\end{equation}
where $\eta=\frac{P_u\beta_0G_0\rho\pi}{N_0W}$. The corresponding GT rate is given by ${R}_{\mathrm{MAC}}(r)=\frac{1}{K_s'}\log_2\left(1+\gamma_{\mathrm{MAC}}(r)\right)$ bps/Hz, which can be shown to decrease as $H$ or $\Theta$ increases, since the decrease in $\frac{1}{K_s'}$ is more significant than the increase in $\log_2\left(1+\gamma_{\mathrm{MAC}}(r)\right)$. The total rate of all GTs in $\mathcal{A}_i'$ is given by
\begin{align}\label{R_UL}
&\tilde{R}_{\mathrm{MAC}}(H,\Theta)=\int_{0}^{2\pi}\int_0^{H\tan\Theta} \rho {R}_{\mathrm{MAC}}(r)rdr d\theta \nonumber\\
=&\frac{1}{\tan^2\Theta}\bigg[\frac{1}{\cos^2\Theta}\log_{2}\!\left(1+\frac{\eta \sin^2\Theta}{\Theta^2}\right)-\log_{2}\left(1+\frac{\eta \tan^2\Theta}{\Theta^2}\right)\nonumber\\
+&\frac{\eta \tan^2\Theta}{\Theta^2}\log_{2}\left(1+\frac{\Theta^2 \tan^2\Theta}{\Theta^2+\eta \tan^2\Theta}\right)\bigg].
\end{align}

Note that interestingly, $\tilde{R}_{\mathrm{MAC}}(H,\Theta)$ is independent of $H$, since as $H$ grows, the decrease in ${R}_{\mathrm{MAC}}(r)$ equally compensates the increase in $K_s'$ in (\ref{R_UL}). On the other hand, similar to the previous two cases, we will investigate the optimal $\Theta$, $\Theta_{\mathrm{MAC}}^\star$, to maximize $\tilde{R}_{\mathrm{MAC}}(H,\Theta)=\tilde{R}_{\mathrm{MAC}}(\Theta)$ regardless of $H$ via numerical examples in the next section.
\section{Numerical Results}\label{sec_num}
In this section, we provide numerical results to validate our analysis. We consider $\beta_0=1.42\times 10^{-4}$, $W=10$MHz, $P_d=10$dBm, $P_u=-10$dBm, $N_0=-169$dBm/Hz, and $\rho=0.005$ GTs/$\hbox{m}^2$, unless specified otherwise.

\begin{figure}[!htb]
	\centering
	\includegraphics[width=12cm]{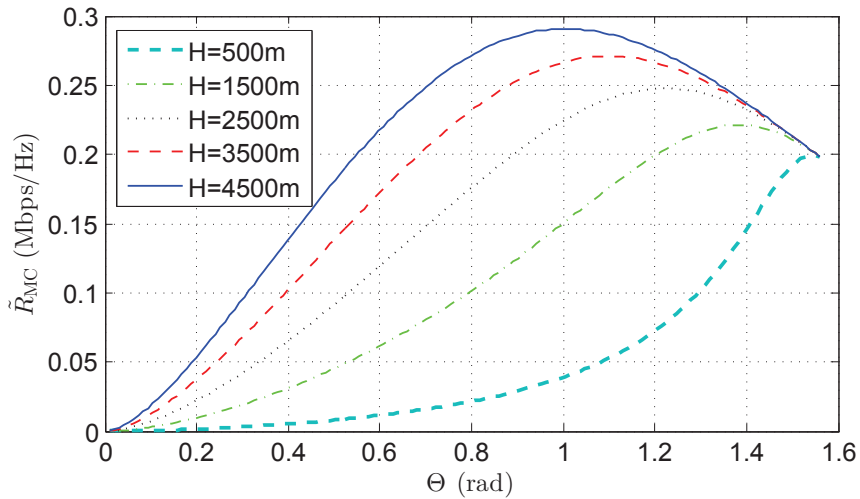}
    \caption{$\tilde{R}_{\mathrm{MC}}$ versus $\Theta$ under different $H$ values in downlink MC.}\label{multicast}
\end{figure}

First, we consider the case of downlink MC and plot $\tilde{R}_{\mathrm{MC}}$ given in (\ref{RMC}) versus $\Theta$ under different values of $H$ in Fig. \ref{multicast}. It can be observed that for any given $\Theta$, $\tilde{R}_{\mathrm{MC}}$ increases with $H$, thus validating Proposition \ref{prop_MC_H}. Moreover, it is observed that given any $H$, $\tilde{R}_{\mathrm{MC}}$ first increases and then decreases as $\Theta$ increases from $0$ to $\frac{\pi}{2}$; while the optimal $\Theta_{\mathrm{MC}}^\star$ that maximizes $\tilde{R}_{\mathrm{MC}}$ is shown to be non-increasing with $H$, which is consistent with our analysis that $\underset{H\rightarrow\infty}{\lim}\Theta_{\mathrm{MC}}^\star\!=\!\Theta_{\min}$.

\begin{figure}[!htb]
  \centering
  \subfigure[$\Theta=\frac{\pi}{10}$ rad]{
  \includegraphics[width=12cm]{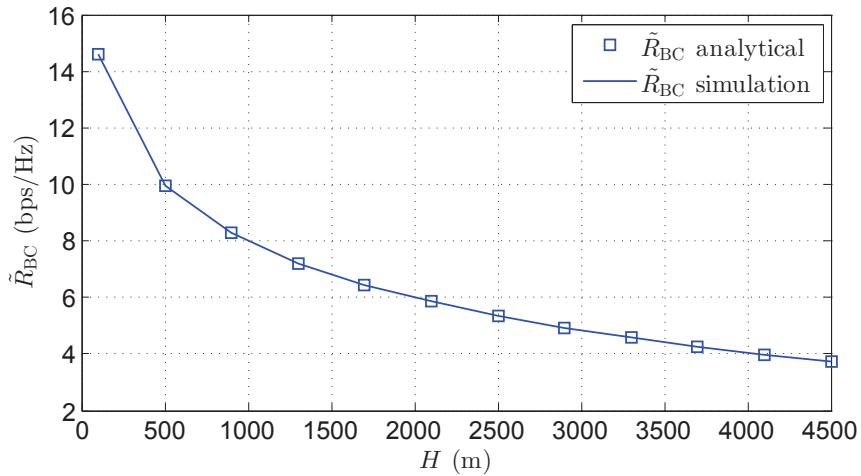}}
  \subfigure[$H=500$m]{
  \includegraphics[width=12cm]{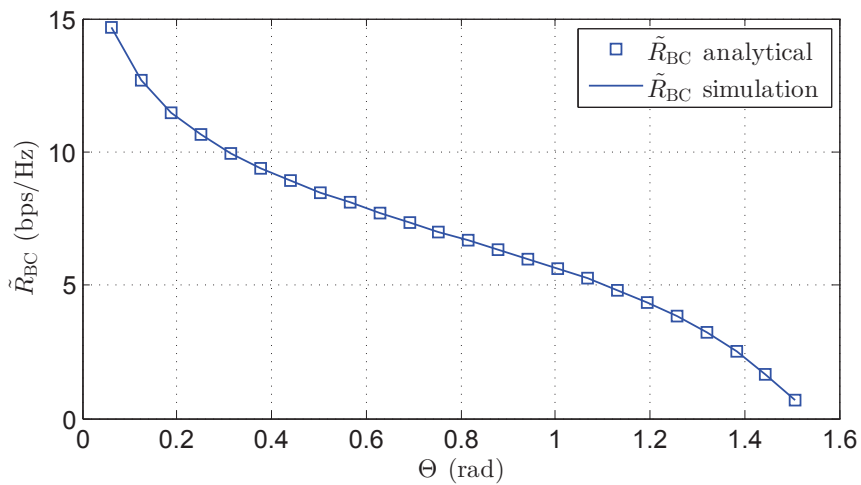}}
  \caption{$\tilde{R}_{\mathrm{BC}}$ versus $H$ or $\Theta$ in downlink BC.
  }
  \label{broadcast}
\end{figure}

Next, we consider the case of downlink BC. In Fig. \ref{broadcast}, we plot the analytical results of $\tilde{R}_{\mathrm{BC}}$ given in (\ref{R_BC-E}), versus $H$ with fixed $\Theta=\frac{\pi}{10}$ rad or versus $\Theta$ with fixed $H=500$m. We also plot the simulation results of $\tilde{R}_{\mathrm{BC}}$ obtained by averaging over $100$ independent realizations of GT locations, which are observed to match closely with the analytical results in Section \ref{sec_broadcast}. It is also observed that $\tilde{R}_{\mathrm{BC}}$ decreases with $H$ as well as $\Theta$, which is consistent with our analysis in Section \ref{sec_broadcast} and suggests that smaller $H$ or $\Theta$ is desirable for maximizing $\tilde{R}_{\mathrm{BC}}$.

\begin{figure}[!htb]
	\centering
	\includegraphics[width=12cm]{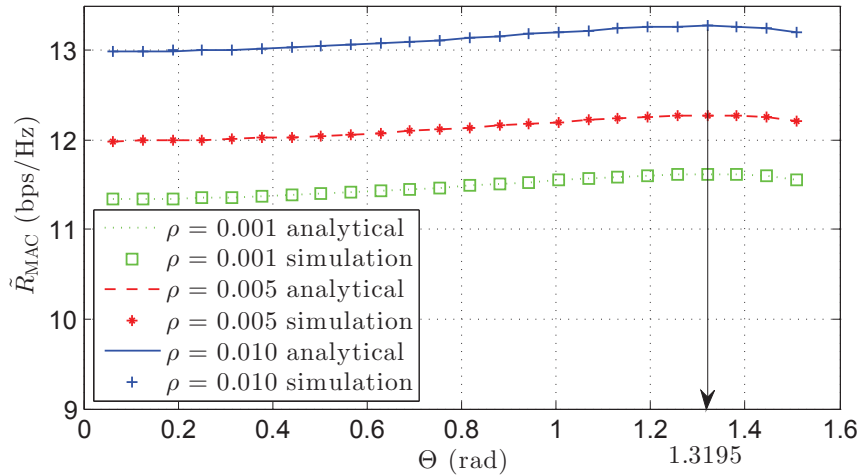}
    \caption{$\tilde{R}_{\mathrm{MAC}}$ versus $\Theta$ under various GT density $\rho$ in uplink MAC.
    }
	\label{uplink}
\end{figure}

Last, we consider the case of uplink MAC. In Fig. \ref{uplink}, we plot both the analytical and simulation results for $\tilde{R}_{\mathrm{MAC}}$ given in (\ref{R_UL}) versus $\Theta$ with arbitrary $H$ and $\rho=0.001, 0.005$ or $0.01$ GTs/$\mbox{m}^2$. It is observed that the analytical and simulation results match well. Moreover, for each given $\rho$, $\tilde{R}_{\mathrm{MAC}}$ first increases and then decreases as $\Theta$ increases from $0$ to $\frac{\pi}{2}$, and the optimal $\Theta^\star_{\mathrm{MAC}}$ that maximizes $\tilde{R}_{\mathrm{MAC}}$ is almost the same for different values of $\rho$, which is $1.3195$ (or $79.7271$ in degree) if $1.3195\in [\Theta_{\min},\Theta_{\max}]$, as indicated in Fig. \ref{uplink}.
\section{Conclusions}
In this letter, we studied the joint altitude and beamwidth optimization problem for UAV-enabled multiuser communication systems. Three fundamental models were studied based on our proposed fly-hover-and-communicate protocol. Our results show drastically different rules for setting optimal altitude and beamwidth values in different multiuser models. We hope that the results provide new and helpful insights for the design of practical UAV-enabled communication systems. Extension of our results to the case with fading UAV-GT channels and/or multiple UAVs is an interesting direction of future work.
\bibliographystyle{IEEEtran}
\bibliography{IEEEabrv,IEEEfull}
\end{document}